# Development and optimization of low power non-thermal plasma jet operational parameters for treating dyes and emerging contaminants

Deepchandra Joshi[1], G. Veda Prakash[2], Shaikh Ziauddin Ahammad[1], Satyananda Kar[2], T R Sreekrishnan[1]

[1]Department of Biochemical Engineering and Biotechnology, [2]Department of Energy Science and Engineering

Indian Institute of Technology Delhi, NewDelhi-110016, India

**Abstract:**

Emerging contaminants (ECs) have come out as the latest class of environmental contaminants, which are highly recalcitrant and toxic in nature. Currently, no suitable rectification methods are available against the ECs, resulting in a continuous increase in their concentration. Non-thermal plasma, as an advanced oxidation process, has been emerging as a promising technology against the ECs treatment. In the present work, a detailed experimental study is carried out to evaluate the efficacy of a non-thermal plasma jet with two dyes, Rhodamine B and Methylene Blue, as model contaminants. The plasma jet provided a complete dye decoloration in 30 min with an applied voltage of 6.5 kV. •OH, having the highest oxidation potential, acts as the main reactive species, which with direct action on contaminants also acts indirectly by getting converted into $H_2O_2$ and $O_3$. Further, the effect of critical operational parameters viz., sample pH, applied voltage (4.5-6.5 kV), conductivity (5-20 $mScm^{-1}$), and sample distance on plasma treatment efficacy was also examined. Out of all the assessed parameters, the applied voltage and sample conductivity was found to be the most significant operating parameter. A high voltage and low conductivity were found to favor the dye decoloration, while the pH effect was not that significant. To understand the influence of plasma discharge gas on treatment efficacy, all the experiments are conducted with Argon and Helium gases under the fixed geometrical configuration. Both the gases provided a similar dye decoloration efficiency. The DBD plasma system with complete dye removal also rendered maximum mineralization of 73 % for Rd. B, and 60 % for Met. Blue. Finally, the system's efficiency against the actual ECs (four pharmaceutical compounds, viz., metformin, atenolol, acetaminophen, and ranitidine) and microbial contaminant (*Escherichia coli*) was also tested. The system showed effectivity in the complete removal of targeted pharmaceuticals and a log 2.5 *E. coli* reduction. The present systematic characterization of dye degradation could be of interest to large communities working towards commercializing plasma treatment systems.

**Abbreviations:**

| | |
|---|---|
| **ECs** | **Emerging contaminants** |
| Rd. B | Rhodamine B |
| Met. Blue | Methylene Blue |
| DBD | Dielectric barrier discharge |
| •OH | Hydroxyl radical |
| OES | Optical emission spectroscopy |
| RONS | Reactive oxygen and nitrogen species |
| WWTs | Wastewater treatment system |
| SBR | Sequencing batch reactor |

## 1. Introduction:

Emerging contaminants (ECs) are the latest class of environmental contaminants which have been recently detected in the environment. The reason for their increased environmental concentration is mainly due to the increased use of chemicals and unresponsible anthropogenic activities [1]. The ECs mainly found in urban wastewater include pharmaceuticals, personal care compounds, endocrine disrupting compounds, pesticides, and antibiotic-resistant bacteria [2]. These contaminants are generally found in a very narrow concentration range of a few $ngL^{-1}$. However, in such low concentrations also, they are extremely harmful to human and aquatic health [3]. The conventional wastewater treatment systems (WWTs) were not designed considering ECs removal; therefore, they show limited efficiency against them [2]. This necessitates advanced treatment systems, which are highly effective in taking care of ECs. The membrane filtration, adsorption process, nanotechnology, and chemical oxidation process are effective advanced processes. However, they have a major limitation that either they are a phase transfer method or require toxic chemicals [4]. These issues limit their commercial use against the ECs. The advanced oxidation process (AOPs), which are based on the high oxidation potential of •OH radicals, has been emerged as a highly effective treatment process [5–8]. Moreover, like other advanced treatment technologies, AOPs don't have major limitations when used on a large scale [9]. The •OH is highly reactive and have the highest oxidation potential among the common disinfectants: •OH (E° = 2.80 V) > $O_3$ (E° = 2.07 V) > $H_2O_2$ (E° = 1.8 V) > $Cl_2$ (E° = 1.36 V) Therefore the •OH can oxidize and degrade organic contaminants most easily and in a non-specific manner. The conventional AOPs mainly include Fenton and Fenton-like reactions, UV/$H_2O_2$/$O_3$ based processes, and photocatalysis [9]. Recently the cold plasma has been found to be an effective AOPs and is already under extensive laboratory trial for finding its large-scale applications in medicine, industrial and wastewater treatment [10,11]. Plasma is the mixture of charged electronic species that could be used to produce various reactive oxygen and nitrogen species (RONS) such as •OH, NO, •$O_2^-$, $O_3$, $N_2^+$, $NO_3^-$, $NO_2^-$ etc. In addition, it also provides other action mechanisms such as UV radiation, heat energy, sonication, etc. [12–14]. All these multiple action mechanisms act synergistically and enhance plasma treatment efficiency. In addition, the plasma operation is also energy efficient, easy to operate, and eco-friendly. Multiple types of plasma systems are reported in the literature, such as pulsed discharge plasma reactor, gliding arc discharge reactor, contact glow discharge electrolysis plasma reactor, atmospheric pressure glow discharge (APGD) plasma reactor, and dielectric barrier discharge (DBD) reactor, etc., [15–19] each of these have their own advantages and setbacks [20]. Although several studies have been reported on dye treatment with non-thermal plasma, the systematic characterization in view of treating ECs are scarce. Thus, more systematic studies are needed to assess the plasma efficacy used for the treatment of wastewater contaminants. Finding the effect of different gases under the same experimental configuration is undoubtedly interesting, considering the future commercialization of the method.

In this work, cold atmospheric pressure plasma jet system is developed, and its feasibility study is done using the Rhodamine B (Rd. B) and Methylene Blue (Met. Blue) dyes. The major emphasis of the present study is to understand the efficacy of developed plasma jets in the treatment of ECs. For this, a systematic study is performed to obtain optimized parameters using dye as a model contaminant. The selected dyes are common textile water pollutants with reported carcinogenic and mutagenic effects and have a strong chemical structure matching the ECs [21,22]. The concentration of dyes was selected according to the ECs concentration range present in sewage water. Another reason for choosing the dyes is due to their simple spectrophotometric analysis. All the experiments were performed using two different plasma gases (Argon and Helium) under the same reactor configuration to evaluate the effect of plasma discharge gas on treatment efficiency. The effects of other parameters such as sample pH, conductivity, applied voltage, plasma- sample distance were also verified and optimized. Physicochemical characterization of the plasma was performed, and the main plasma reactive species were quantified in the liquid phase. The reactive species analysis helped in understanding the plasma action mechanism.

After understanding the reaction mechanism and attaining the system's optimized conditions, treatment of actual ECs was performed. For this, the removal of four pharmaceuticals viz., metformin ($C_4H_{11}N_5$), atenolol ($C_{14}H_{22}N_2O_3$), acetaminophen ($C_8H_9NO_2$), and ranitidine ($C_{19}H_{30}BiN_4O_{10}S$) from secondary treated sewage water was performed. The removal of targeted pharmaceuticals in sequencing batch reactor (SBR) and SBR plasma-based hybrid reactor was also evaluated for comparative analysis. On the other hand, the antibiotic resistance bacteria (ARB) are an important class of microbial ECs whose degradation is also very difficult. ARB has a vital role in the development and proliferation of antibiotic resistance, which has been declared the biggest problem of the 21$^{st}$ century by WHO [23]. Therefore, the system's efficiency is also verified against the microbial contaminant removal, and for this, *Escherichia coli* (*E. coli*) was selected as model ARB, and its treatment was also done in the same optimized condition used for the pharmaceuticals.

## 2. Experimental section

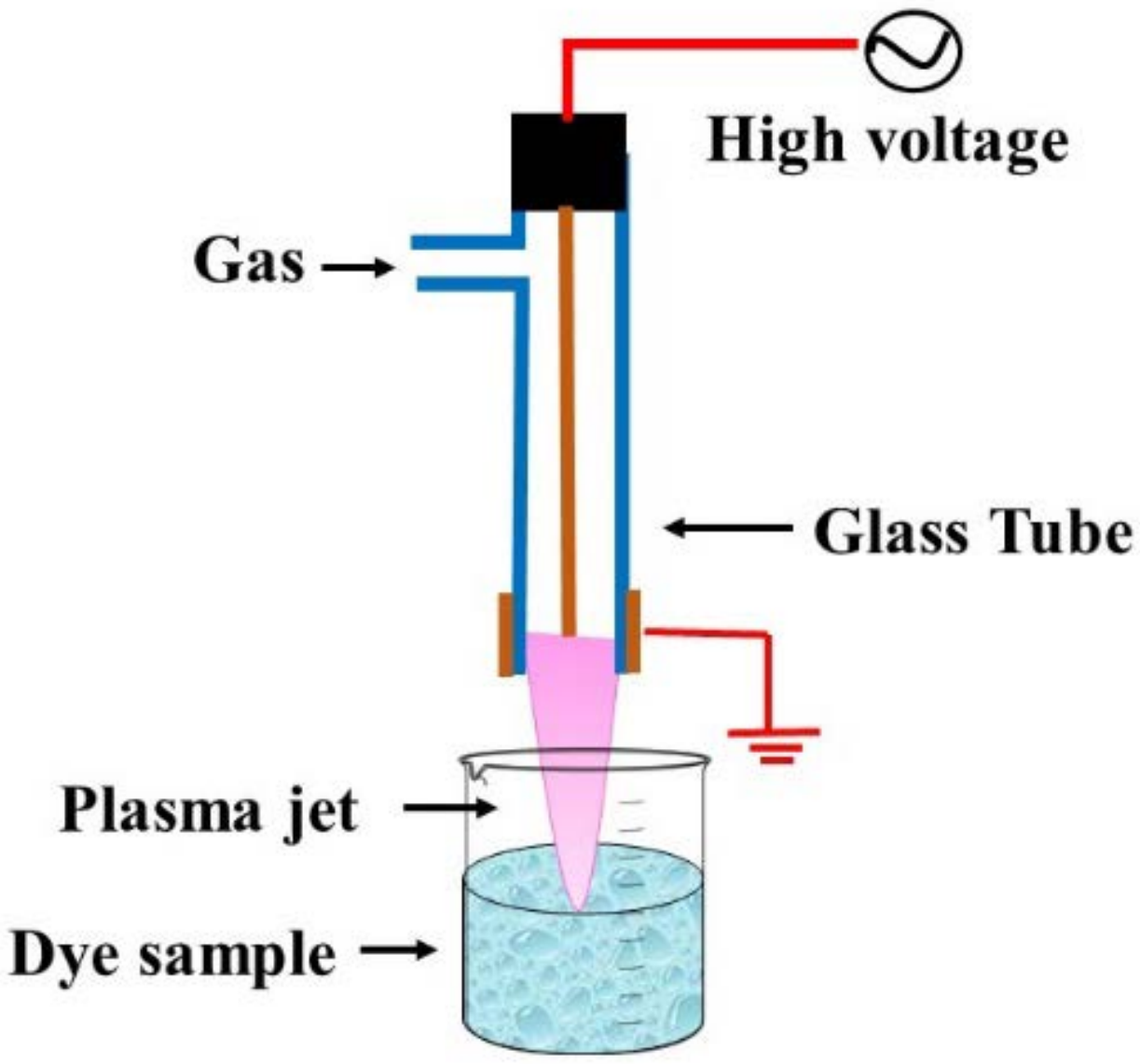

**Fig. 1 Schematic of the atmospheric pressure cold plasma jet treatment system**

**2.1 Experimental setup**

The schematic of the experimental setup is shown in Fig. 1. The plasma jet system is a combination of a high voltage sinusoidal power supply and a plasma reactor. The indigenously developed high voltage sinusoidal AC power supply of 0 - 6.5 kV, 25 kHz was used to generate the plasma. The plasma reactor was made up of a glass tube with an inner and outer diameter of 4 mm and 6 mm, respectively. A 1.5 mm diameter copper rod was used as a high voltage electrode, and a copper wire of 1 mm diameter was wrapped around the glass tube and used as a ground electrode. Electrodes are arranged concentrically, which make a cross-field configuration of the plasma reactor. Argon and Helium gases were used independently for the generation of plasma. The position of the electrodes and other plasma reactor configurations was fixed for the entire experiment. Under fixed geometrical conditions, with the applied voltage of 6.5 kV, 25 kHz frequency, and gas flow of 3 LPM for Argon and 4 LPM for Helium, a plasma jet length of ~ 3 cm was obtained in the ambient air with both the gases. Typical voltage and current profiles for argon plasma jet are shown in Fig. S1. The single jet DBD plasma system used in this study can be scaled up following linear scale-up criteria. Multiple arrays of this plasma jet and appropriate power supply can be used to develop a larger system with increased capacity.

**2.2 Materials and Instrument**

The dye and all other chemicals used in this study are of analytical grade and were procured from Sigma-Aldrich, India. The media for microbial analysis was from HiMedia. The colored dye solutions were prepared in RO water (conductivity ~0.1 mScm$^{-1}$), and for fluorescent dye, Milli Q (conductivity ~0.05 µScm$^{-1}$), water was used. The commercially available compressed Argon and Helium gas (99.99%) was used as the plasma discharge gas. The spectrophotometer (JENWAY 7305), pH meter (EUTECH instruments), conductivity meter (Hanna Instruments), Fluorescence meter (Spectra plus 384), Ion chromatography (Metrohm), and TOC analyzer (SHIMADZU) were used for performing the appropriate analysis. The applied voltage was measured using a Tektronix high voltage probe P6015A. The emission profile from the plasma discharge was acquired by the Ocean

Optics spectrometer HR-4000 model connected with the optical fiber cable (200 µm) at a 5 mm distance radially near the glass nozzle.

### 2.3 Experimental procedure

The colored dye samples were prepared in RO water, and initially, a stock concentration of 100 ppm was developed. A 1 ppm sample was used for plasma treatment from the stock solution. Experiments were performed with an operating voltage of 6.5 kV (5 W average power) and 25 kHz sinusoidal voltage. The dye feasibility study was performed with the DBD plasma jet system, and then the other parameters such as pH, conductivity, input voltage, and plasma sample distance were optimized to obtain a complete dye decoloration. All the experiments were performed with fixed applied voltage until specifically mentioned. For the pH experiments, the sample's initial pH was adjusted using the HCl and NaOH, and for the conductivity experiment, NaCl was used to adjust the sample's initial conductivity. All the experiments were done in triplicate and then averaged. The decoloration efficiency was calculated using equation 1 after measuring the absorbance of pre-and post-treated samples. The Rd. B and Met. Blue shows a maximum UV visible absorption at a wavelength of 554 and 663 nm, respectively.

$$\textbf{Decoloration efficiency} = \frac{\text{initial concentration } (C_0) - \text{Final concentration } (C_t)}{\text{initial concentration } (C_0)} \times 100 \qquad \textbf{(1)}$$

For the kinetic study, the results obtained were fitted in the first-order rate equation (equation 2), from which the apparent rate constant was also calculated.

$$\ln \frac{\text{initial concentration } (C_0)}{\text{Final concentration } (C_t)} = kt \qquad \textbf{(2)}$$

To determine the extent of dye mineralization under the specific treatment conditions, total organic carbon (TOC) analysis was performed. Finally, to verify the treatment efficiency of the plasma system against the actual ECs, 4 pharmaceuticals (metformin, atenolol, acetaminophen, and ranitidine) and a microbial contaminant (*E. coli*) were treated in the optimized condition obtained from the dye study.

### 2.4 Reactive species determination

A variety of short-lived (few nanoseconds) and long-lived (more than 1 s) reactive oxygen and nitrogen species (RONS) gets produced during plasma action at the gas-liquid interface, and the chemical reactions involved in their formation are shown in equation 5-17 [24]. The high voltage plasma generates the free electrons ($e^-$) and other excited species, which then ionize the interacting water molecule. This starts a chain reaction, in which the ionized water molecule may break down to form the hydronium ion, free H atom, and •OH radical. The preference of these interfacial species in air or liquid can be known from their dimensionless Henry constant ($K_H$) value. The molecule with a $K_H$ value greater than one has more preference in the air compared to liquid. The $K_H$ value of important reactive species formed in plasma is •OH - $6.92\times10^2$, $H_2O_2$ - $1.92\times10^{6,}$ and $O_3$ - $1\times10^{-1}$ [25]. The short-lived •OH initiate the formation of long-lived reactive species. Therefore, identifying the long-lived species further affirms the existence of the short-lived species. The bulk diffusion of these reactive species is also crucial as it acts as a rate-limiting step. Those plasma species that interact with the liquid compounds have a faster diffusion than non-reactive species. While, the non-reactive species diffusion occurs simply by the absorption process [26]. To understand the mechanism of plasma interaction with contaminants, the analysis of these reactive species is essential. In those, •OH, is one of the main reactive species, and its quantification was done using the coumarin-based chemical probe method. It is an indirect method in which the coumarin, a non-fluorescent dye, reacts with •OH, to form a fluorescent compound 7-hydroxy coumarin as per the mechanism shown in (supplementary Fig. S2) [27]. The fluorescence intensity of 7 hydroxy coumarins is directly proportional to the •OH concentration [28]. Coumarin gives a maximum UV- visible absorption at 277 nm, and 7 Hydroxy coumarin, gives the maximum fluorescence at 455 nm after initially exciting at 332 nm [28]. 1 mM aqueous coumarin sample was used to determine the aqueous •OH concentration. In the plasma alone, •OH presence was analyzed using optical emission spectroscopy (OES). The $H_2O_2$ is another essential and long-lived reactive species, and its quantification was done using the Potassium Iodide (KI) based colorimetric method given by Ovenston [29]. In this method, the $H_2O_2$ reacts with KI in the presence of Ammonium molybdate ($(NH_4)_6Mo_7O_{24}$), and Iodine ($I_2$) is liberated, which makes the solution yellow-colored (equation 3).

$$\mathbf{H_2O_2 + 2I' \rightarrow 2OH' + I_2} \quad \mathbf{(3)}$$

The intensity of the developed color is directly proportional to the $H_2O_2$ concentration, and by measuring the absorbance of color at 353 nm and comparing it with a standard $H_2O_2$ curve, the $H_2O_2$ concentration in unknown samples can be determined. Ozone ($O_3$) analysis was done using the commercially available $O_3$ test kit procured from the Hanna instrument, which is based on the Diethyl-p-phenylene diamine (DPD) method. The Nitrogen species ($NO_3^-$ and $NO_2^-$) and other ions/nutrients formations were quantified using ion chromatography.

### 2.5 ECs analysis

Liquid chromatography-mass spectroscopy (LCMS) was used for the pharmaceuticals analysis. The difference in pre-and post-treated samples concentration on a percentage basis was used to evaluate the removal efficiency.

*Microbial sample:* An overnight grown *E. coli* cell culture in nutrient broth was centrifuged at 10,000 rpm for 10 min. The obtained cell pellet was then resuspended in 10 mL PBS buffer. The resuspended solution in triplicate was used as the samples for plasma treatment. The treated sample was then serially diluted using PBS buffer, and suitable dilution was plated on the nutrient agar plate to observe the effect of plasma action on *E. coli* deactivation. For comparison, the non-treated and control samples were also plated in a similar manner. The *E. coli* inoculated in PBS buffer acted as the positive control, and PBS buffer without inoculation was used as a negative control. The plates were incubated overnight at 35 ± 2 °C. The difference in log $CFUmL^{-1}$ calculated using equation 4 provided the system's microbial inactivation efficiency.

$$\boldsymbol{E.coli}\ \mathbf{deactivation} = \frac{\mathbf{C_i - C_f}}{\mathbf{C_i}} \quad \mathbf{(4)}$$

Where $C_i$ = initial microbial colony count in log $CFUmL^{-1}$ and $C_f$ = microbial colony count in log $CFUmL^{-1}$ after plasma treatment.

## 3. Results and discussion

### 3.1 RONS quantification

The •OH formation during plasma action has been shown in Fig. 2 (a & b). The figure shows an increase in the sample's fluorescence with increased plasma treatment time due to the conversion of coumarin into 7 Hydroxy coumarin. This conversion decreases the coumarin concentration in the sample, which can be seen from the decrease in the sample's absorption spectrum shown in Fig. S3 (a & b). In Fig. 2 (a & b), an unusual behavior (instead of continuous increase, the fluorescence intensity starts decreasing after a particular time) can be seen after 5 min (with Argon plasma) and 10 min (with Helium plasma). This behavior is observed because the plasma action becomes so prominent that it also starts degrading the 7 hydroxy coumarin molecules after this specific time. Therefore, the sample's fluorescence intensity decreases. Using a higher concentration dye sample can be a solution, but with coumarin, it is not possible due to its solubility limitation of around 1 mM in the aqueous solution. The conversion of •OH into other reactive species also occurs, and this also limits the available number of •OH molecules getting trapped by coumarin to form 7 hydroxy coumarin. This limits the detection of an actual number of •OH molecules, resulting in a difference in the observed and actual •OH concentration after some time. This problem is also present with other chemical probes, and previously, Tampieri et al. used Terephthalate (TPA) as a probe for •OH analysis and have reported that the TPA formation also deviates from linearity after a certain time [30]. Therefore, for maintaining the uniformity in data, the •OH and other plasma species analysis was performed for up to 20 min only. The •OH formation in the plasma jet was detected using OES, and the obtained spectra have been shown in Fig. S4 for Argon plasma and Fig. S5 for Helium plasma. The characteristic peak at 309 nm (prominent in the •OH band of 306-312 nm) in the OES spectrum is of the •OH, and the nearby other peaks are of other RONS. The •OH concentration in the liquid sample (RO water and neutral pH) in a treatment time of 1-20 min is shown in Table 1. The maximum estimated •OH concentration with Argon and Helium plasma is 14.04 µM and 15.04 µM, respectively. The concentration obtained is in correlation with other similar studies [13,31]. The study performed in the reference [32], have used 20 W input power and in 10 min the trend of •OH concentration obtained with different plasma gases was $O_2$ (1.19 mM) > Air (1.03 mM) > $N_2$ (0.64 mM) > Ar (0.59 mM). Further, Kovačević, et al., compared five different gases (Ar, He, Air, Oxygen, and Nitrogen) for •OH formation in a DBD reactor in direct contact with water. Highest •OH concentration was obtained with oxygen as the plasma feed gas, while a comparable concentration gets produced with the Ar and He gases [33]. Due to difficulty in •OH detection, many studies have also used the fluorescence value or the extent of contaminant degradation to express the •OH concentration [34,35]. In the present study, both the Argon and Helium gases provided almost a similar concentration of •OH. In such a case, Argon gas is more advantageous as it is a heavier gas, which helps it penetrate more into the liquid and promote the reactive species diffusion in the bulk liquid. Still, we used both gases in all our dye studies to evaluate the difference in treatment efficiency. The •OH with a half-life of 1 µs and diffusion coefficient, $D_g$ of $2.0 \times 10^{-9}$ $ms^{-2}$ can only travel a maximum distance of 10 µm [32]. The diffusivity of •OH in liquid gets increased by its conversion into $H_2O_2$ (equation 8), which has more stability and can diffuse deeply inside the liquid [36]. Then as per equations 12 and 13, again •OH gets formed by the breakdown of $H_2O_2$. It is reported that 99.6 % of total $H_2O_2$ formed in the plasma system is by the recombination of •OH [25]. Therefore, $H_2O_2$ concentration can also be used as a measure to find the plasma system's efficiency in producing •OH [37,38]. Table 1 shows the $H_2O_2$ concentration obtained in this study, and it shows an increase in $H_2O_2$ concentration with increasing treatment time. In 20 mins treatment time, we have obtained 562 µM and 448 µM of $H_2O_2$ with the Argon and Helium plasma, respectively. With 5 W of average applied power, the $H_2O_2$ concentration obtained in the present experiments is in a similar range to that obtained by other researchers, such as 680 µM in 10 min with an input power of 20 W [32]. Whereas higher concentrations of $H_2O_2$ viz.,1765 µM and 1176 µM with 20 min treatment time and 150 W, 100 W applied power, respectively were obtained [38]. In the present work the $H_2O_2$ concentration obtained was very high compared to •OH because before trapping the •OH molecule by coumarin, it gets converted into $H_2O_2$ or any other reactive species. Due to the interconversion of reactive species, a major portion of the generated $H_2O_2$ also gets lost by decomposition to form the $H_2O$ and $O_2$ (equation 11) [32]. Ozone ($O_3$) is another important stable reactive species formed during the plasma action (equation 15). The ozone formation was also evaluated in the RO water, and similar to $H_2O_2$, a continuous increase in its concentration is obtained. A maximum concentration of 320 µM and 111 µM in 20 min of plasma treatment was obtained with the Argon and Helium plasma, respectively (Table 1). Multiple studies have reported that the formation of $O_3$ in plasma reactors takes place mainly when $O_2$ or air has been used as the plasma feed gas [33,39]. In a study by Kinandana et al., it was also found that $O_3$ formation has also been getting significantly affected by the system geometry. In their studies, the researchers has compared the series and parallel DBD configurations and reported that the series configuration was better as it provided 2000 µM more $O_3$ concentration than the parallel configuration [40]. With argon and helium as the plasma feed gas, we have got a concentration of 320 µM and 111 µM, respectively (table 1). Without oxygen, the $O_3$ in the plasma system gets produced from the conversion of •OH. The series of reactions involving this conversion process has been represented in the equation 5-17. In a study by Teng et al., the influence of $O_3$ on the formation of $NO_2^-$ and $NO_3^-$ was studied. In this study, the researchers have used a different ratios of $O_2$: $N_2$ in plasma feed gas [41] and found a positive effect of $O_3$ on nitrogen species, especially $NO_3^-$. Whereas in the present experiments, we have obtained a low $O_3$ concentration, and also, there was no direct source of nitrogen. Therefore, very less $NO_3^-$ and $NO_2^-$ formation occurred, and it was beyond the detectable range of ion chromatography. In literature also, $NO_3^-$ and $NO_2^-$ formation was mainly reported when nitrogen gas or air was used as the plasma feed gas [33,42,43]. The low nitrogen formation by our system provides an additional advantage, as these nitrogen species themselves are an important water contaminant [44]. In the study by Meropoulis et al., it has been concluded that these nitrogen species do not have much role in pollutant degradation and cause a decrease in sample pH during plasma action [45]. Similar findings have also been reported in other studies [33]. Due to the formation of these nitrogen species in undetectable concentrations, no significant pH change in post treated sample was observed in our case. After 30 min of plasma treatment time, the sample pH changed from neutral to slightly acidic (pH 6) which was due to the formation of organic acids formed from the dye degradation by-products [46]. The concentration of these organic acids is also very less, due to the systems high mineralization efficiency, about which a detailed description has been provided in the section 3.10. Finally, after analyzing the results of reactive species formation and understating their chemistry with the contaminants [9,34,43], it is apparent that the •OH and $H_2O_2$ are the main reactive species responsible for the dye degradation during plasma action.

$$2H_2O + e^- \rightarrow H_3O^+ + OH^- + e^- \quad (5)$$

$$Ar^* + H_2O \rightarrow H^+ + \bullet OH + e^- \quad (6)$$

$$e^- + H_2O \rightarrow \bullet OH + H\bullet \quad (7)$$

$$OH\bullet + OH\bullet \rightarrow H_2O_2 \quad (8)$$

$$\bullet OH + H_2O_2 \rightarrow H_2O + HO\bullet_2 \quad (9)$$

$$HO\bullet_2 + H_2O_2 \rightarrow O_2 + H_2O + OH\bullet \quad (10)$$

$$2H_2O_2 \rightarrow 2H_2O + O_2 \quad (11)$$

$$HO\bullet_2 + \bullet OH \rightarrow H_2O + O_2 \quad (12)$$

$$*e + H_2O_2 \rightarrow H\bullet + \bullet OH + e^- \quad (13)$$

$$O_2 + e^- \rightarrow e^- + 2\bullet O \quad (14)$$

$$\bullet O + O_2 \rightarrow O_3 \quad (15)$$

$$O_3 \rightarrow \bullet O + O_2 \quad (16)$$

$$O_3 + H_2O_2 \rightarrow \bullet OH + H_2O_2 + O_2 \quad (17)$$

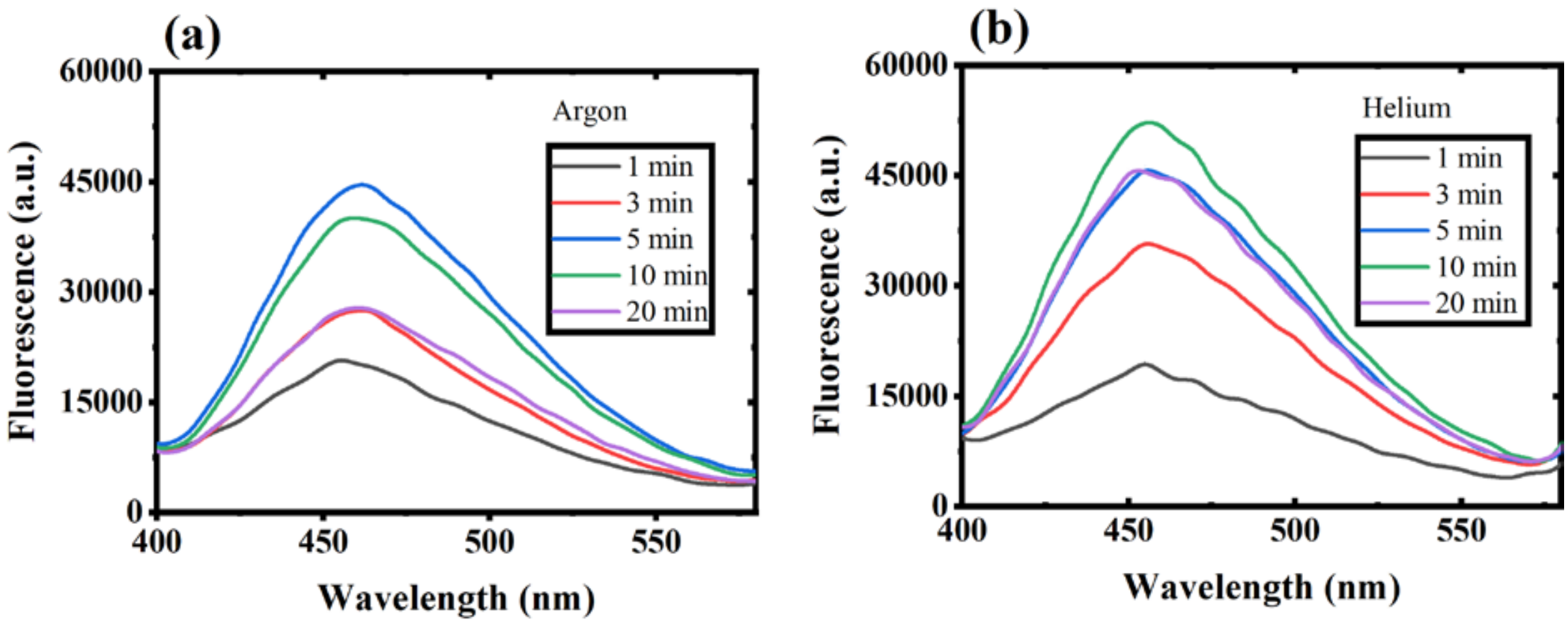

**Fig. 2 The fluorescence spectrum of 1mM coumarin sample treated with (a) Argon and (b) Helium DBD plasma. (Voltage- 6.5 kV)**

**Table 1: Formation of RONS in DBD plasma. (Sample DI water, voltage- 6.5 kV)**

| Time (min) | •OH (μM) | | $H_2O_2$ (μM) | | $O_3$ (μM) | |
|---|---|---|---|---|---|---|
| | Argon | Helium | Argon | Helium | Argon | Helium |
| 1 | 6.01 | 5.00 | 195.37 | 88.84 | 8.34 | 2.08 |
| 3 | 8.04 | 9.74 | 245.34 | 157.25 | 16.67 | 4.17 |
| 5 | 14.04 | 12.93 | 408.67 | 211.27 | 41.67 | 8.34 |
| 10 | 12.55 | 15.04 | 470.34 | 258.32 | 277.9 | 55.63 |
| 20 | 8.56 | 13.13 | 562 | 448.73 | 320 | 111.25 |

### 3.2 Effect of pH

In any of the AOPs, the sample pH is considered to influence the treatment efficiency. To evaluate the effect of pH on dye decoloration, the samples with acidic (pH 4), neutral (pH 7), and basic (pH 10) pH were treated. The results have been shown in Fig. 3 (a-d). The data represents a minor or almost negligible effect of pH on the degradation of the model dyes. Specifically, samples treated with Helium have shown marginally higher decolorization in a shorter treatment time (in 5 min > 85 %), whereas it took a little longer for Argon.

Overall, the acidic pH was found to slightly favor the dye decoloration process. The acidic pH samples of Rd. B and Met. Blue showed 100 % removal, while complete color removal was not obtained in the basic and neutral pH samples. A higher contaminant degradation in an acidic pH was also reported by other researchers [47,48]. While the basic pH favors those reactions in which the important RONS gets decomposed. Such as the reaction between the •OH and $H_2O_2$ in which hydroperoxyl ($HO•_2$) radicals get formed (equation 9), which has less oxidation potential than •OH [49]. The •OH radical has 100 times faster reactivity ($k = 7.1 \times 10^9\ M^{-1}s^{-1}$) with $HO•_2$ than $H_2O$, and in this reaction, few •OH molecules get consumed (equation 12). The self-decomposition of $H_2O_2$ is also favored in basic pH according to equations 9 and 11. In basic pH, the $OH^-$, $CO_3^-$, $HCO_3^-$, etc., ions are present, and they also act as the •OH scavenger. With this in basic pH, there is an easy conversion of •OH into its conjugate base $O^{\cdot-}$. The •OH acts as a electrophile while reacting with organic molecules, and the $O^{\cdot-}$ is a nucleophile. Therefore, these two radicals can form different intermediate products, leading to different degradation pathways [50]. Further, it has been reported that plasma treatment shifts the treated sample pH towards the acidic side due to nitric/ nitrous acid formation. These acids are formed from the nitrogen species ($NO_2^-$ and $NO_3^-$), mainly generated when air or nitrogen is used as the plasma feed gas [51,52]. A very less pH variation has been observed when nitrogen was not present in the plasma feed gas [42,53]. Bolouki et al. had compared the pH variation in the plasma-treated sample when argon, oxygen, nitrogen, and air were used as the plasma feed gas [53]. They have reported no significant pH change with argon and oxygen, while pH decreased sharply with air and nitrogen plasma. In the same study, it was also found that the nitrogen species also acts as the •OH quencher, which was evidenced by absence of •OH peak in air and nitrogen plasma OES spectra. As we have also used argon and helium as the plasma feed gas; therefore, not much pH drop in the treated sample was observed.

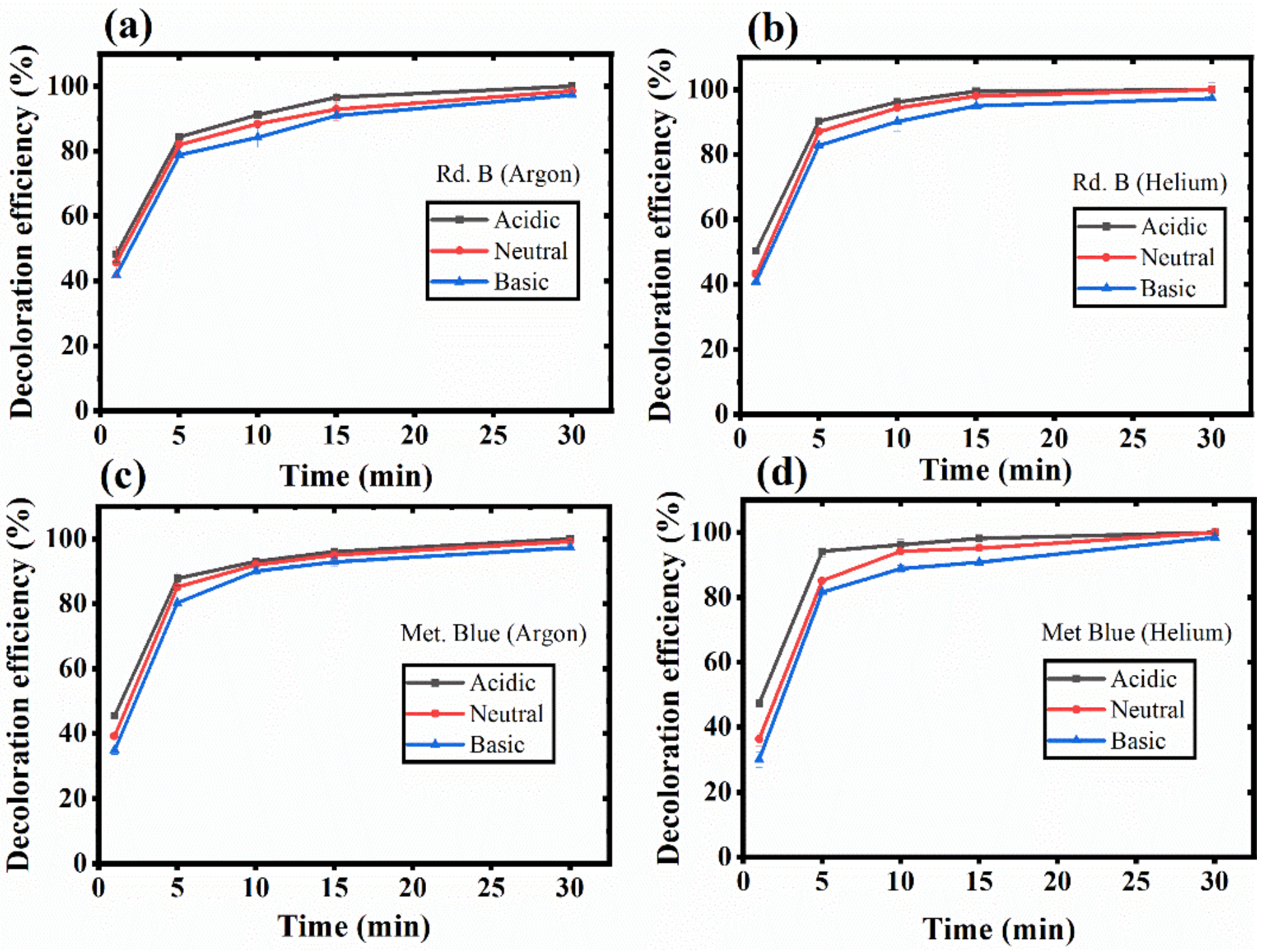

**Fig. 3 Dye decoloration at different pH values with increasing treatment time. (a) Rd. B-Argon (b) Rd. B-Helium (c) Met. Blue- Argon and (d) Met. Blue- Helium DBD plasma.**

### 3.3 Effect of conductivity

The sample conductivity strongly affects the plasma discharge characteristics, which further influences the reactive species generation. The conductivity becomes more critical when the discharge is applied directly over the liquid solution. In this case, the plasma behavior such as discharge channel, current, length, and propagation depends on the sample conductivity [54].

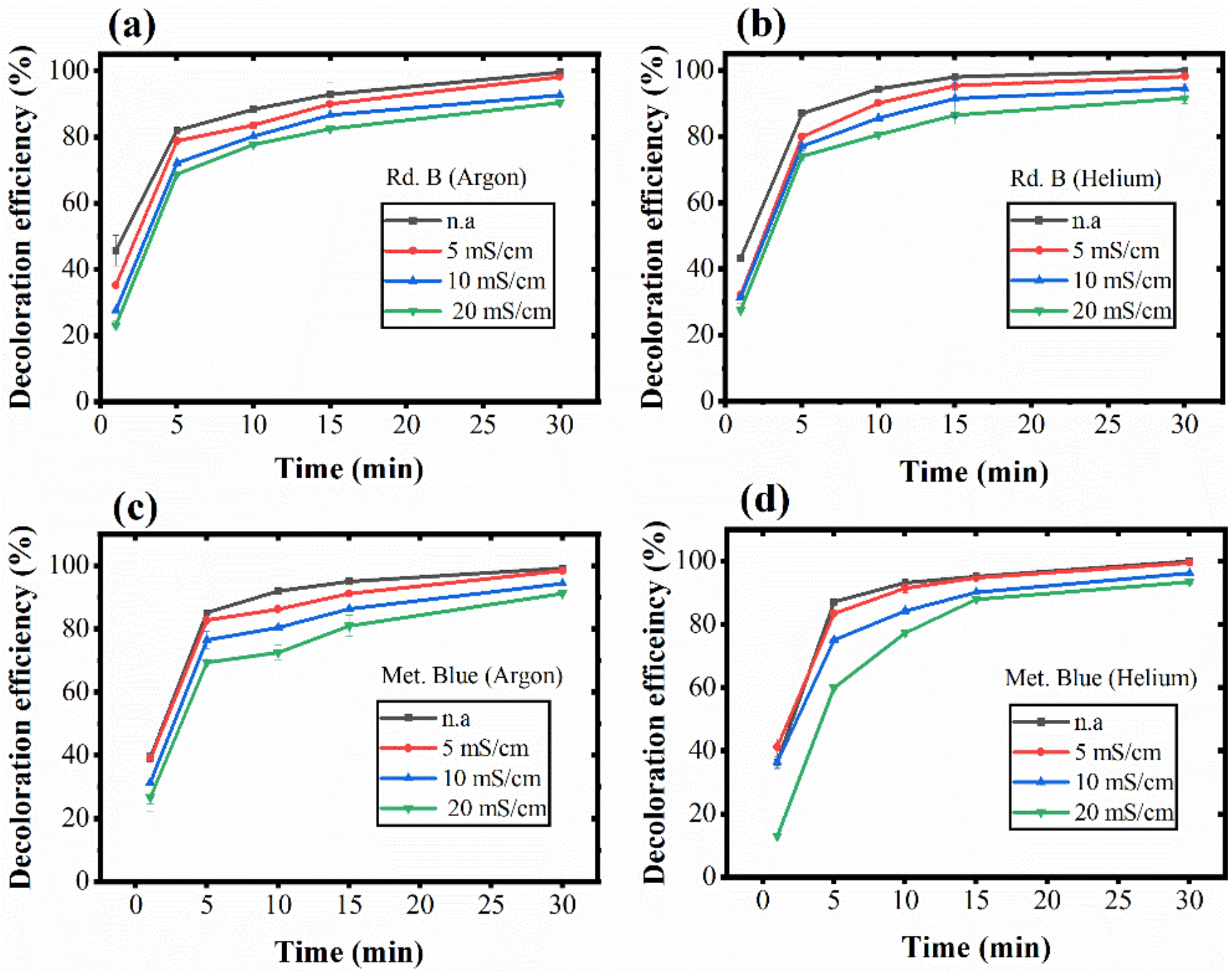

**Fig. 4 effect of sample's initial conductivity on dye decoloration with increasing treatment time. (a) Rd. B-Argon (b) Rd. B-Helium (c) Met. Blue- Argon and (d) Met. Blue- Helium DBD plasma.**

Our study target is to check the DBD cold plasma feasibility for wastewater treatment. The wastewater generally has concentrated salts and inorganics, increasing its conductivity to the range of tens of $mScm^{-1}$. Therefore, the conductivity experiment was done at a neutral pH and in the conductivity range of non-altered (n.a), 5, 10, and 20 $mScm^{-1}$. The n.a sample has a conductivity value of ~0.16 $mScm^{-1}$ for Rd. B and ~0.28 $mScm^{-1}$ for Met. Blue. The results have been shown in Fig.4 (a-d), which depicts that for the first 5 min, a high dye removal was observed with both the dyes and the gases, and from 5 to 15 min the dye decoloration increased gradually. After this, almost a saturation is obtained, and for non-altered (n.a) and 5 $mScm^{-1}$ conductivity samples, complete dye decoloration was obtained in 30 min. In the same duration with 10 and 20 $mScm^{-1}$ conductivity samples, dye decoloration was obtained only up to 90 %. This shows a decrease in treatment efficiency with increasing conductivity. Such behavior was observed because the plasma discharge becomes more adverse at higher conductivity, resulting in plasma conversion to spark discharge with an increment in plasma intensity and discharge current for both Argon and Helium gases. Similar observations were reported in other studies where the decrease in reactive species formation was noticed with an increase in intensity and discharge current as the sample conductivity increased [43,55]. This reduction in reactive species could be the possible reason for reduced treatment efficiency in the present case. An opposite behavior is also reported in a study where a much higher conductivity sample (up to 100 mS/cm) was used [48]. The high conductivity was favorable in their case because the other plasma effect (ozone and UV) became more prominent, which ultimately improved the plasma efficiency, but with a high energy requirement.

### 3.4 Effect of applied voltage

The applied voltage directly influences the formation of high-energy electrons, which results in more reactive species generation. Therefore, the decoloration efficiency should increase with an increase in applied voltage. Therefore, the present system's potential in different applied voltage value was evaluated. The chosen voltage value was 4.5, 5.5 and 6.5 kV and their corresponding average power value was 3.1 ,3.9 and 4.9 W respectively. The results have been shown in Fig. 5 (a-d), which shows the complete removal of Rd. B and Met. Blue in 30 min with 6.5 kV applied voltage, while it was only 91% for Rd. B with Argon, and 93% for Met. Blue with Argon at 4.5 kV applied voltage. With lower power to achieve the same level of removal, a longer time is needed. This shows that the plasma depends on the applied voltage, which is an indirect measure of applied power. In higher power, more dense charged electron species are generated with increased physical effects resulting in increasing the efficiency of the treatment process [54].

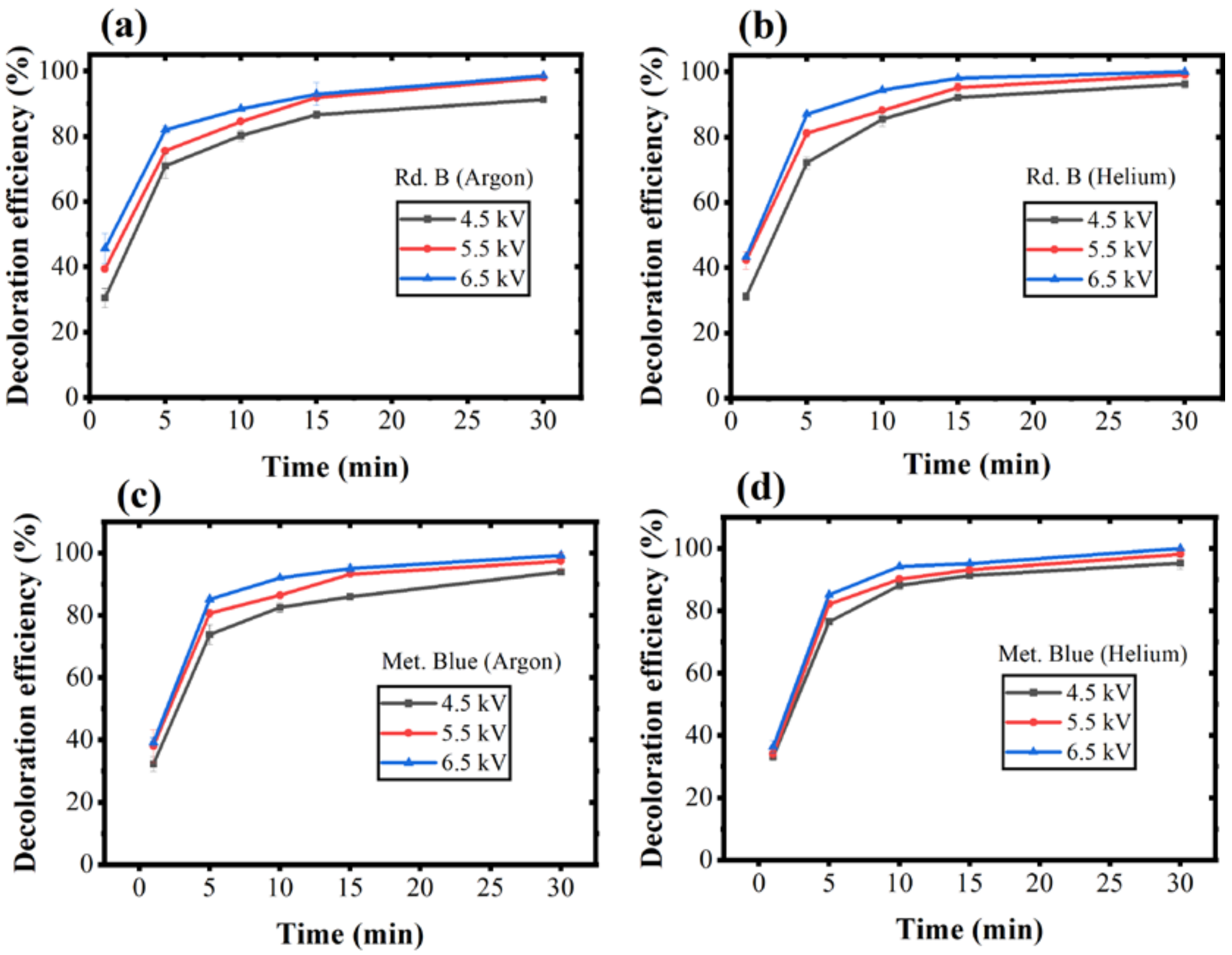

**Fig. 5 Applied voltage effect on dye decoloration with increasing treatment time. (a) Rd. B-Argon (b) Rd. B-Helium (c) Met. Blue- Argon and (d) Met. Blue- Helium DBD plasma.**

### 3.5 Effect of sample distance from the plasma

With the increase in plasma sample distance, the gaseous phase short-lived RONS gets decayed before they get diffused inside the liquid [30]. Therefore, finding an optimum distance help to obtain a point at which the maximum plasma penetration in the liquid could be obtained. The highest removal was obtained when the sample was placed at 2 cm, i.e., at the middle of the plasma plume. At 1 cm distance (top of plasma plume), the dye decoloration gets decreases, while at 3 cm distance (tip of plasma), the least removal is observed. Few reactive species are present at the plasma tip, and plasma penetration efficiency is also not strong enough if the sample plasma distance is more. While in a much closer distance, the coming discharge gas gets bounced back after hitting the water layer, as of which suitable plasma plume formation didn't take place. In addition to this, narrower plasma-sample distance fluctuation in the liquid surface causes improper Tylor cone formation, resulting in instability in the top layers of liquid, which also affects the formation of plasma discharge [51,56]. Therefore, keeping the sample at an optimum distance, which in our case was around the middle of the plasma plume, would provide maximum efficiency.

### 3.6 Kinetic study

The dye samples were treated at a time interval of 0 to 30 min to find the effect of increasing treatment time on the dye decoloration process. For both the dyes, during the initial 5 min, the dye decoloration was exponential, which becomes gradual up to 15 min. After this, till complete removal, almost a saturating behavior was observed until complete removal, as shown in Fig. 6 (a & b). A similar decoloration pattern of both the dye was obtained with both the Argon and Helium gas. The 90% removal of both the dyes was obtained in 20 min of plasma treatment, but it took an extra 10 min for complete dye decoloration. The extra time was needed to degrade chromophores and resistive intermediates released from the breakdown of dye. The sample, in an extended treatment time with direct point treatment of liquid results in mild heating of the samples. Due to this, a slight loss in liquid volume due to evaporation was observed. In this study, during 30 min of treatment, the maximum temperature of the sample reached was around 50°C.

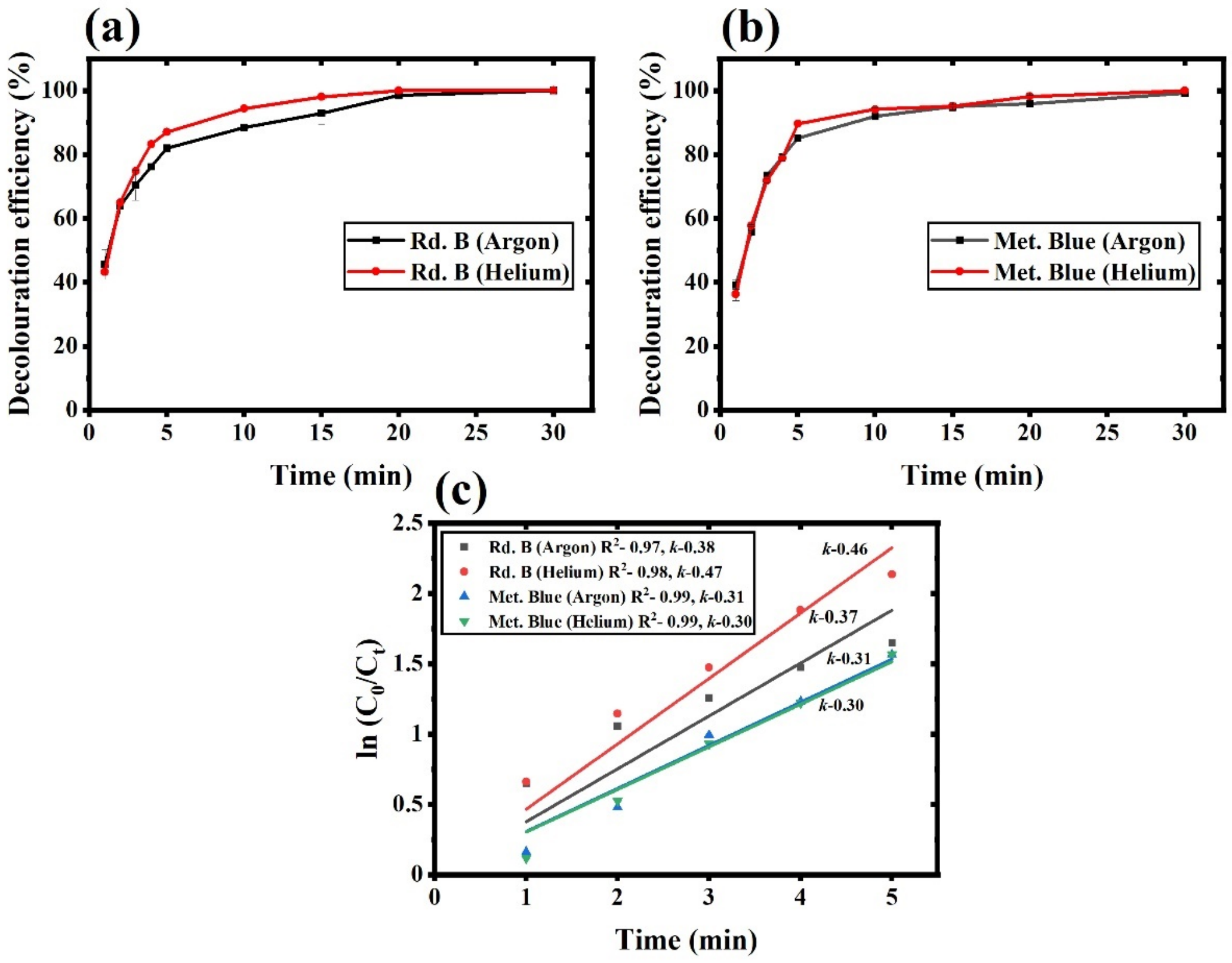

**Fig. 6 Kinetic study of dye decoloration (a) for Rd. B, (b) for Met. Blue and (c) logarithmic representation of ($C_0/C_t$) vs time graph for determining kinetic constants of Argon plasma-treated Rd. B sample.**

**Table 2: Kinetic parameters of the dye degradation process for the initial 5 min and complete dye decoloration process.**

| | Up to 5 min | | | | Overall | | | | |
|---|---|---|---|---|---|---|---|---|---|
| Sample | Decoloration (%) | $k$ (min$^{-1}$) | $t_{(1/2)}$ (min) | $R^2$ (%) | Decoloration (%) | Mineralization (%) | $k$ (min$^{-1}$) | $t_{(1/2)}$ (min) | $R^2$ (%) |
| Rd. B (Ar) | 82 | 0.46 ± 0.03 | 1.82 | 0.97 | 100 | 73 | 0.12 ± 0.02 | 5.78 | 0.91 |
| Rd. B (He) | 87 | 0.37 ± 0.02 | 1.47 | 0.99 | 100 | 45 | 0.19 ± 0.03 | 3.65 | 0.87 |
| Met. Blue (Ar) | 85 | 0.31 ± 0.01 | 2.24 | 0.99 | 99 | 60 | 0.17 ± 0.02 | 4.08 | 0.95 |
| Met. Blue (He) | 89 | 0.30 ± 0.01 | 2.31 | 0.99 | 100 | 35 | 0.18 ± 0.02 | 3.85 | 0.94 |

The ln ($C_0/C_t$) vs. time graph was plotted, which has been shown in Fig. 6 (c), and the apparent or observed rate constant *k* was calculated, using which the half-life of the dye molecule during plasma treatment was calculated. The dye decoloration results were consistent with the first-order rate equation, and therefore the rate constant depends on the dye concentration at any given time. Table 2 lists the rate constant for the initial 5 mins and for the overall dye decoloration reaction. In all cases, the rate constant for the initial 5 min is more than the overall reaction, which is also reflected in the $t_{1/2}$ value of the dye molecule. Initially, breakdown of the complete dye molecule occurs, but as the process continues, resistive intermediates get formed, due to which the dye decoloration rate decreases. In the extended time, the dye decoloration also losses its linearity, and the reaction gets shifted from first order to pseudo-first-order rate equation.

**3.8 Effect of dye concentration**

To evaluate the efficiency of plasma on a different level of contamination, samples with an increasing dye concentration were treated. The result of increasing dye concentration has been depicted in Fig. S6, from which it can be concluded that increasing dye concentration limits the plasma treatment efficiency. As in all cases, the maximum dye decoloration occurred during the initial 5 min of treatment. Therefore, for this experiment, sample treatment was done only for 5 min. Around 80 % removal was observed during the selected treatment time in the 1, 5, and 10 ppm samples with both Argon and Helium gas. In the 50-ppm sample, the removal gets lowered by 50 %, and an average removal of 40 % was observed. The majority of the ECs in wastewater was found to be in the range of a few ppm; therefore, plasma treatment will be effective against them. If, in some specific type of wastewater, a higher concentration of ECs will be present, then by increasing the treatment time or input voltage of the DBD plasma system, a high treatment efficiency can be achieved.

**3.9 Individual effect of RONS species**

It is reported that certain molecules with higher reactivity with RONS act as the scavenger molecule and consume the RONS before they could oxidize the contaminants [57,58]. Methanol is such a scavenger molecule for •OH, and its presence would decrease the plasma effectivity. Here we used this property of methanol for evaluating the individual effect of •OH on the reactor's treatment efficiency [59]. For this, 10 % methanol acting as the •OH radical quencher was added to the dye sample, and dye decoloration was compared with the sample without any added methanol. Like this, the individual effect of other main reactive species was also determined. To find out the individual effect of $H_2O_2$, the samples were treated with externally added $H_2O_2$. A major limitation in determining the individual effect of $H_2O_2$ during plasma treatment is maintaining the $H_2O_2$ concentration at a constant level. Due to continuous formation of $H_2O_2$, its concentration gets changed. Still, to get rough idea on the effect of $H_2O_2$, the highest reported concentration of $H_2O_2$ obtained during plasma treatment in current study was used as the sole treating reagent in a batch reactor. In the batch reactor only $H_2O_2$ in a concentration of 600 μM was added to the dye samples and treatment was performed for 5 min. The results obtained from this experiment was compared with the experiment conducted using plasma for the same treatment time. The results obtained from these two studies have been compared in Fig. S7. In the sample where the •OH was quenched, the least dye decoloration was observed, 42.45 % (Rd. B- Ar) and 44.21 % (Met. Blue- He). While the $H_2O_2$ treated, samples have shown a decoloration of 68.33 % (Rd. B) and 66.95 % (Met. Blue). In Fig. S8 the results of dye decoloration with lowest to highest obtained $H_2O_2$ concentration (50 to 600 μM) could also be found out. As, it can be seen that in •OH quenched samples least dye decoloration was obtained, which affirms that •OH has the highest contribution in plasma treatment. The •OH with a direct attack on dye molecule also has a role in the formation of other reactive species.

### 3.10 Dye mineralization and Presumption of degradation mechanism

Mineralization is the efficiency of a treatment process to convert the pollutant into mineral acids (HCl, Nitric acids etc.), $H_2O$, and $CO_2$. In the absence of proper mineralization, the degradation by-products having similar or higher toxicity can get formed. Therefore, the decoloration experiment was followed by the mineralization experiment to assess systems mineralization efficiency, and it was expressed in terms of total organic carbon (TOC). The decrease in the TOC value shows a decrease in the effluent toxicity level and the formation of mineralized products. The results of the mineralization experiment have been shown in table 2, which shows that Argon as plasma discharge gas provided more mineralization 73 % (Rd. B-Ar), 60 % (Met. Blue- Ar), compared to Helium with 45 % (Rd. B- He), and 35 % (Met. Blue- He). Compared to the Argon plasma, low mineralization was achieved with the Helium plasma. This shows that although the Helium has high dye decoloration efficiency, but its mineralization efficiency is low compared to Argon. The mineralization efficiency obtained in this study was higher than other similar studies performed on dye treatment [60]. The results obtained show that the present treatment system with high decoloration can also provide good dye mineralization. The degradation pathway by which the mineralization of Rd. B [61] and Met. Blue [4,62] occurs has been well explained in the respective works of literature.

### 3.11 Pharmaceuticals and Microbial contamination removal

Table 3: The removal of the pharmaceutical compound in a conventional method and DBD plasma treatment system. (SBR treatment time- 24 hr, and volume- 1 L)

| **Compound** | **Inlet (SBR) ($pgmL^{-1}$)** | **Outlet (SBR) ($pgmL^{-1}$)** | **Plasma ($pgmL^{-1}$)** | **SBR + plasma ($pgmL^{-1}$)** |
|---|---|---|---|---|
| Metformin | 18674 | 965 | 888.4 | 18.4 |
| Atenolol | 349 | 3.8 | 4.02 | ND |
| Acetaminophen | 4.7 | 1.71 | 0.17 | ND |
| Ranitidine | 150 | 14.41 | 0.88 | ND |

The detailed dye decoloration study showed that both the plasma discharge gases provided similar efficiency under the same reactor configuration. Compared to Helium, the Argon plasma provided a higher dye mineralization efficacy, and it is also a relatively cheaper gas. Therefore, Argon gas was selected to perform further studies. The system's efficiency against the actual ECs present in wastewater was determined by treating the sequencing batch reactor (SBR) treated secondary wastewater and targeting the removal of 4 pharmaceuticals. Efficient removal of all the selected pharmaceuticals was observed, and the results are shown in Table 3. The table also includes the comparative results of the SBR, plasma, and SBR- hybrid plasma system. The SBR has a working volume of 1 L, a cycle time of 24 hr, and 10% sludge, and it shows a high removal because of the contaminants bioadsorption by the sludge. The 5 min Argon plasma treatment also showed a high removal of pharmaceuticals except for the Metformin. Almost a complete removal is observed when the plasma-hybrid treatment system was used. This shows that a combination of plasma and conventional wastewater treatment systems is an effective ECs treatment method.

Finally, the plasma system efficiency was also evaluated with the microbial contaminants considering *E. coli* as model antibiotic resistance bacteria (ARB). The *E. coli* was treated using both the Argon and Helium plasma, and Argon plasma was found to be more effective with log 2.5 *E. coli* deactivation in 6 min of plasma treatment (Fig. 7a). The typical image of Argon plasma-treated sample with increasing treatment time has also been shown in Fig. 7b, and the absence of *E. coli* colony can be seen in 6 min treated sample plate. Cell wall degradation is an important mechanism of microbial deactivation by plasma action [25, 52]. The *E. coli* (Gram-negative) has a thin cell wall (<10 nm), and to check plasma action on other cell wall types *Bacillus subtilis*, a Gram-positive bacteria with a thick cell wall (20-80 nm), was also treated under the same treatment condition [64]. The plasma-treated samples show a log removal of 2.45 for *E. coli* and 2.42 for *B. subtilis* (Fig. 7 c&d). Therefore, a similar removal was observed with both the bacterial types, and it can be concluded that plasma acts non-specifically and is equally effective against different bacterial types.

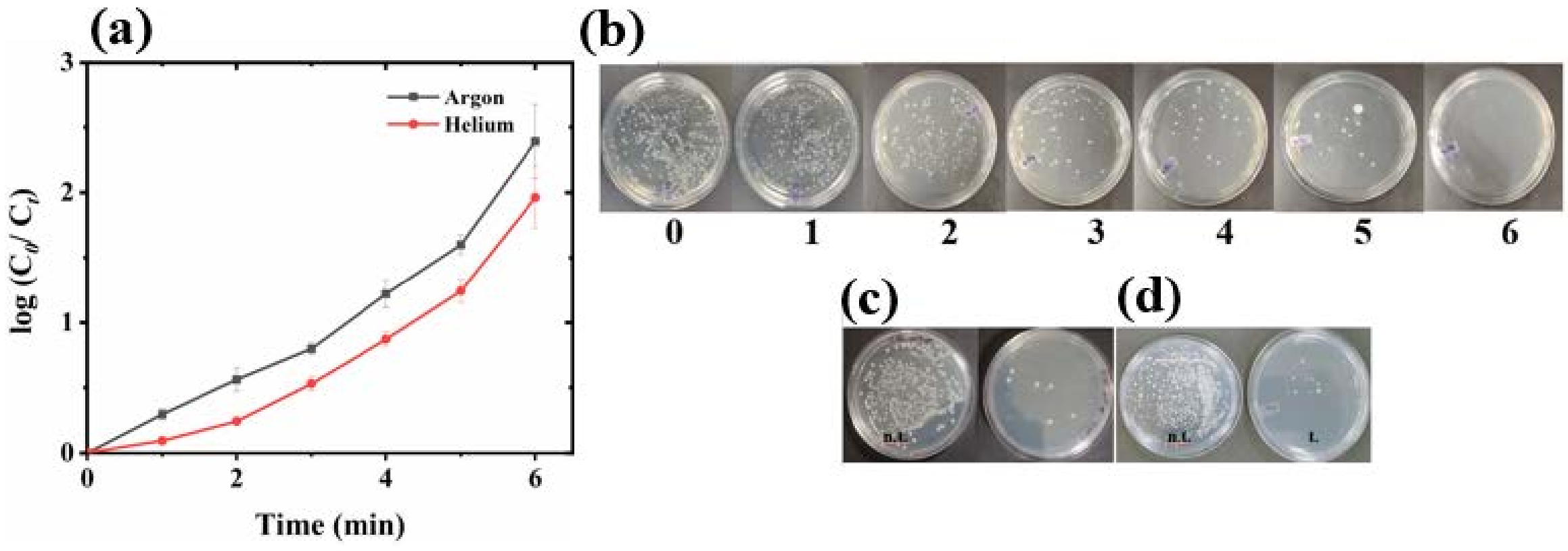

**Fig. 7 Microbial treatment with plasma (a) The change in microbial concentration with time, (b) media plates showing the removal of microbial colonies with increasing treatment time and pre and post-treated (c) *E. coli and* (d) *B. subtilis*. ($C_0$- log 8 CFU $mL^{-1}$)**

## 4. Conclusion

In this study, the DBD plasma jet system assessment against the ECs treatment was performed using Rd. B and Met. Blue as model contaminants. The system's operational parameters were optimized using the dye as the model contaminant, the fastest dye decoloration up to 80 % in all cases occurred during the first 5 min, and a complete dye decoloration was achieved within 30 min of treatment. By the analysis of RONS species, the formation of •OH, $H_2O_2$ and $O_3$ were confirmed. With further detailed analysis, it was also found that the •OH and $H_2O_2$ acts as the main action mechanism during plasma action. The system was also efficient in providing high dye mineralization which is extremely useful in ensuring the absence of any toxicity in the effluent. A comparative study of Argon and Helium as plasma discharge gas was performed, and it was observed that both gases provides a similar treatment efficiency. Considering the treatment efficiency and cost-effectivity, the utilization of Argon gas is suggested. Finally, under the optimized conditions, the plasma system was highly effective against the plasma treatment of sewage water (consisting of four pharmaceutical chemicals) and microbial ECs. The pharmaceutical removal was also checked in SBR (conventional treatment system) and SBR plasma hybrid reactor for the comparative study. The complete removal of the pharmaceuticals was observed in a plasma SBR hybrid reactor. These results show the efficacy of a developed non-thermal plasma systems for the treatment of ECs. Moreover, the present systematic study of plasma system operational parameters and optimization for dye decoloration will aid in establishing the technology for large-scale ECs treatment.

## Conflict of interest

The authors declare no conflict of interest.

## Acknowledgment

This work was supported by grants from the IIT Delhi FIRP program grant number - MI02081.